\documentclass{jpp}
\usepackage{graphicx,bm,color}
\graphicspath{{./fig/}{./png/}}
\usepackage{amsmath,url}
\usepackage{amssymb}

\newcommand{\EQ}{\begin{equation}}
\newcommand{\EN}{\end{equation}}
\newcommand{\EQA}{\begin{eqnarray}}
\newcommand{\ENA}{\end{eqnarray}}
\newcommand{\eq}[1]{(\ref{#1})}

\newcommand{\Eq}[1]{equation~(\ref{#1})}
\newcommand{\Eqs}[2]{equations~(\ref{#1}) and~(\ref{#2})}

\newcommand{\Sec}[1]{section~\ref{#1}}

\newcommand{\Fig}[1]{figure~\ref{#1}}

\newcommand{\Tab}[1]{table~\ref{#1}}

\newcommand{\bra}[1]{\langle #1\rangle}

{}
{}

{}
{}

{}
{}
{}
{}
{}
{}
{}
{}
{}
{}
{}
{}
{}
{}

{}

\newcommand{\hatk}{\hat{k}}
\newcommand{\hatkk}{\hat{\bm{k}}}

{}

{}
{}
{}

\newcommand{\tildeAA}{\tilde{\mathbf{A}}}

\newcommand{\xxx}{\hat{\mbox{\boldmath $x$}} {}}

\newcommand{\kk}{\bm{k}}

\newcommand{\xx}{\bm{x}}

\newcommand{\rr}{\bm{r}}

\newcommand{\BB}{\bm{B}}

\newcommand{\JJ}{\bm{J}}

\newcommand{\AAA}{\bm{A}}

\newcommand{\UU}{\bm{U}}

\newcommand{\nab}{{\bm{\nabla}}}

\newcommand{\RRRR}{\mbox{\boldmath ${\sf R}$} {}}
\newcommand{\SSSS}{\mbox{\boldmath ${\sf S}$} {}}

\newcommand{\ii}{{\rm i}}

\newcommand{\DD}{{\rm D} {}}

\newcommand{\dd}{{\rm d} {}}

\newcommand{\const}{{\rm const}  {}}

\def\la{\mathrel{\mathchoice {\vcenter{\offinterlineskip\halign{\hfil
$\displaystyle##$\hfil\cr<\cr\sim\cr}}}
{\vcenter{\offinterlineskip\halign{\hfil$\textstyle##$\hfil\cr<\cr\sim\cr}}}
{\vcenter{\offinterlineskip\halign{\hfil$\scriptstyle##$\hfil\cr<\cr\sim\cr}}}
{\vcenter{\offinterlineskip\halign{\hfil$\scriptscriptstyle##$\hfil\cr<\cr\sim\cr}}}}}

\def\Sp{\mbox{\rm Sp}}
\def\Ma{\mbox{\rm Ma}}

\def\Lu{\mbox{\rm Lu}}

\def\EEM{{\cal E}_{\rm M}}

\def\EM{E_{\rm M}}

\def\cs{c_{\rm s}}

\def\xiM{\xi_{\rm M}}

\def\vA{v_{\rm A}}

\def\vAz{v_{\rm A0}}
\def\vAm{v_{\rm Am}}

\def\HM{H_{\rm M}}
\def\EM{E_{\rm M}}

\def\Brms{B_{\rm rms}}

\def\urms{u_{\rm rms}}

\newcommand{\s}{\,{\rm s}}

\newcommand{\cm}{\,{\rm cm}}

\title[Turbulent decay controlled by two conserved quantities]{Turbulent magnetic decay controlled by two conserved quantities}

\author[A. Brandenburg and A. Banerjee]{
Axel Brandenburg\aff{1,2,3,4}\corresp{\email{brandenb@nordita.org}}
\, \and \,
Aikya Banerjee\aff{5}
}
\affiliation{
\aff{1}Nordita, KTH Royal Institute of Technology and Stockholm University,
Hannes Alfv\'ens v\"ag 12, SE-10691 Stockholm, Sweden
\aff{2}The Oskar Klein Centre, Department of Astronomy,
Stockholm University, AlbaNova, SE-10691 Stockholm, Sweden
\aff{3}McWilliams Center for Cosmology \& Department of Physics,
Carnegie Mellon University, Pittsburgh, PA 15213, USA
\aff{4}School of Natural Sciences and Medicine, Ilia State University,
3-5 Cholokashvili Avenue, 0194 Tbilisi, Georgia
\aff{5}Department of Physical Sciences, Indian Institute of Science Education and Research Kolkata, Mohanpur 741246, West Bengal, India
}

\date{\today}

\begin{document}

\maketitle

\begin{abstract}
The decay of a turbulent magnetic field is slower with helicity than without.
Furthermore, the magnetic correlation length grows faster for a helical
than a nonhelical field.
Both helical and nonhelical decay laws involve conserved quantities:
the mean magnetic helicity density and the Hosking integral.
Using direct numerical simulations in a triply periodic domain, we show
quantitatively that in the fractionally helical case the mean magnetic
energy density and correlation length are approximately given by the
maximum of the values for the purely helical and purely nonhelical cases.
The time of switchover from one to the other decay law can be
obtained on dimensional grounds and is approximately given by
$I_\mathrm{H}^{1/2}I_\mathrm{M}^{-3/2}$, where $I_\mathrm{H}$ is the
Hosking integral and $I_\mathrm{M}$ is the mean magnetic helicity density.
An earlier approach based on the decay time is found to agree with our
new result and suggests that the Hosking integral exceeds naive estimates
by the square of the same resistivity-dependent factor by which also
the turbulent decay time exceeds the Alfv\'en time.
In the presence of an applied magnetic field, the mean magnetic helicity
density is known to be not conserved, and we show that then also the
Hosking integral is not conserved.
\end{abstract}


\section{Introduction}

In recent years, there has been significant interest in the study of
decaying turbulent magnetic fields.
One of the main applications has been to the understanding of the
magnetic field evolution during the radiation-dominated era of the
early universe \citep{BEO96, CHB01, BJ04}.
The special case with finite magnetic helicity has been studied and
understood for a long time \citep{Hatori84, BM99}.
It is the prime example of large-scale magnetic field growth due to an
inverse cascade.
The possibility of such an inverse cascade is explained by the
conservation of magnetic helicity \citep{Frisch+75}.
However, even in the absence of magnetic helicity, an inverse cascade can
develop \citep{Kahn+13,Zrake14,BKT15}, and it is well explained by the conservation of
what is now called the Hosking integral \citep{HS21,HS23,Zhou+22,BSV23},
which is the correlation integral of the magnetic helicity density.

In all the cases mentioned above, either the magnetic helicity density was
vanishing, so the spectral magnetic helicity was zero at all wavenumbers
and the decay governed by the conservation of the Hosking integral,
or the magnetic helicity density was finite and the spectral magnetic
helicity had the same sign at all wavenumbers, so the decay was governed
by the conservation of the mean magnetic helicity density.
A special situation was studied in the work of \cite{BKS23}, where the
magnetic helicity was finite, but it was balanced by fermion chirality
of the opposite sign so that the net chirality was vanishing.
For such a system, the decay was again successfully explained by the
conservation of the Hosking integral, which was adapted to
include the chirality from the fermions.

We have seen that the Hosking integral can be applied to broad
ranges of systems where magnetic helicity is still important locally,
but globally, the net magnetic helicity vanishes.
However, there is an important class of astrophysically relevant systems,
where the magnetic field is not generated by magnetogenesis, as in the
early universe, but by dynamo action.
This means that some of the kinetic energy of turbulent motions is
converted into magnetic energy.
It is important to stress that, even if the velocity field were helical,
i.e., if there is finite kinetic helicity in the system, as is generally
the case when there is rotation and stratification of density and/or
velocity, magnetic helicity conservation still precludes the generation
of magnetic helicity, at least on dynamical timescales \citep{Ji99}.

In the aforementioned helically driven large-scale dynamos, magnetic
helicity can be generated at small scales, but it is then balanced by
magnetic helicity at large scales so as to conserve magnetic helicity.
Alternatively, we can also say that magnetic helicity is produced at
large scales, for example by the tilting of buoyantly rising magnetic
flux tubes in cyclonic convective events, as envisaged by \cite{Par55}.
Magnetic helicity conservation then implies magnetic twist of opposite
sign at smaller scales.
In practice, because there is always finite magnetic diffusivity, which
acts especially on small scales, the magnetic helicity from large scales
will, after some time, dominate the total magnetic helicity owing to the
loss at small scales where the magnetic helicity has the opposite sign.
Therefore, there is always a small imbalance between the contributions
from small and large length scales.
It is therefore a situation that is only partially suited to the
phenomenology involving the conservation of the Hosking integral.

If we now were to turn off the driving, the turbulence would gradually
decay.
This decay should then be governed by the conservation of both the
Hosking integral and the mean magnetic helicity density.
Both helical and nonhelical cases lead to inverse cascading, where
the magnetic field decays more slowly than the velocity field, leading
ultimately to a magnetically dominated state.
Such conditions could apply to the decay of a magnetic field produced
in a proto-neutron star.
There, we expect a turbulent dynamo to occur that is driven by
convection \citep{Thompson+Duncan93}.
This would happen when the neutrino opacity is large enough to prevent
neutrinos from escaping freely \citep{Epstein79, Burrows+Lattimer86}.

Another source of turbulence in proto-neutron stars could be the
magnetorotational instability that results from the radially outward
decreasing angular velocity gradient associated with collapsed
material having an approximately constant angular momentum density
\citep{Guilet22}.
In both cases, the turbulence itself has kinetic helicity of opposite
signs in the northern and southern hemispheres (negative in the north
and positive in the south).
In each hemisphere, this leads to dynamo action of the type described
above, but the magnetic helicities have opposite signs not only in the
two hemispheres, but also on small and large length scales.
One would thus focus only on one hemisphere and ignore the interaction
between north and south.
The magnetic helicities from small and large length scales would then
nearly cancel.
Such fields have been called ``bihelical'' and their decay properties
were first studied by \cite{YB03}.
They found that the positive and negative contributions rapidly mix and
annihilate, and that the ratio of the magnetic helicity spectrum to the
magnetic energy spectrum has local extrema at both small and large scales,
although the latter is dominant in an absolute sense.

Since the net magnetic helicity of a bihelical magnetic field does
not vanish exactly, and since the mean magnetic helicity itself is an
important conserved quantity, we are confronted with a situation where
the magnetic decay is governed by two conserved quantities.
Investigating this aspect in a more controlled fashion is the main
purpose of this paper.

In an earlier paper, \cite{Tevzadze+12} studied a case with
fractional helicity.
They found that the correlation length developed a steeper growth
(indicative of magnetic helicity domination) at a specific moment that
depends on the value of the magnetic helicity as well as on the initial
values of the magnetic energy and the magnetic correlation length.
This consideration provided a quantitative estimate for the time of the
switchover from non-helical to helical scalings.
A similar estimate was provided by \cite{HS21} based on the scaling
of the Hosking integral $I_\mathrm{H}$.
One may then ask whether the time of the switchover from a decay
controlled by $I_\mathrm{H}$ to that controlled by the mean magnetic
helicity density $I_\mathrm{M}$ can be computed based on dimensional
arguments.
Indeed, given that the quantity $I_\mathrm{H}$ has dimensions
$\cm^9\s^{-4}$ and $I_\mathrm{M}$ has dimensions $\cm^3\s^{-2}$
\citep[see][and note that the magnetic field is here understood to be
in Alfv\'en units with dimensions $\cm\s^{-1}$]{Bra23}, a combination
of $I_\mathrm{H}$ and $I_\mathrm{M}$ that yields a time would be
$I_\mathrm{H}^{1/2}I_\mathrm{M}^{-3/2}$.
It will turn out that this is indeed the time of switchover between
the two regimes.

\section{Our model}

\subsection{Basic equations}

We simulate the compressible magnetohydrodynamic equations with
an isothermal equation of state with constant sound speed $\cs$, so the
pressure $p$ and the density $\rho$ are related by $p=\rho\cs^2$.
The equations for the magnetic vector potential, $\AAA$, the velocity
$\UU$, and the logarithmic density $\ln\rho$, are
\begin{equation}
\frac{\partial\AAA}{\partial t}=\UU\times\BB+\eta\nabla^2\AAA,
\label{dAdt}
\end{equation}
\begin{eqnarray}
\frac{\DD\UU}{\DD t}=-\cs^2\nab\ln\rho
+\frac{1}{\rho}\left[\JJ\times\BB+\nab\cdot(2\rho\nu\SSSS)\right],
\label{dUdt}
\end{eqnarray}
\begin{equation}
\frac{\DD\ln\rho}{\DD t}=-\nab\cdot\UU,
\end{equation}
where $\DD/\DD t=\partial/\partial t+\UU\cdot\nab$ is the advective
derivative, $\BB=\nab\times\AAA$ is the magnetic field,
$\JJ=\nab\times\BB/\mu_0$ is the current density,
$\mu_0$ is the vacuum permeability, $\nu$ is the viscosity, and
${\sf S}_{ij}=(\partial_i U_j+\partial_j U_i)/2-\delta_{ij}\nab\cdot\UU/3$
are the components of the traceless rate-of-strain tensor $\SSSS$.
Our computational domain is a periodic cube of size $L^3$, and
$k_1=2\pi/L$ is the smallest wavenumber.
Since the mass in the domain does not change, the volume-averaged density
is constant in time, i.e., $\bra{\rho}=\const\equiv\rho_0$.
Here and below, angle brackets denote volume averaging.

\subsection{Initial conditions}

In our idealized studies, we focus on the decay governed by two
conserved quantities (Hosking integral and mean magnetic helicity density).
We construct an initial magnetic vector potential in Fourier space as
$\tildeAA(k)=\RRRR(k;\varsigma)\tildeAA^\mathrm{nhel}$, where
\begin{equation}
\mathsf{R}_{ij}(\kk;\varsigma)=\delta_{ij}-\hatk_i\hatk_j
+\ii\hatk_\ell\varsigma\epsilon_{ij\ell}
\end{equation}
is a matrix with $\hat{k}_i$ being the components of the unit vector
$\hatkk=\kk/k$, $|\varsigma|\leq1$ is a nondimensional parameter that
quantifies the fractional helicity, and $\tildeAA^\mathrm{nhel}$ is a
nonhelical field with random phases and possesses the desired spectrum
for the magnetic field $\Sp(\BB)=k^2\Sp(\AAA)$, i.e.,
\begin{equation}
\Sp(\BB)=\frac{A_0 k^{\alpha}}{1+(k/k_0)^{5/3+\alpha}},
\label{SpB}
\end{equation}
where $A_0$ is an amplitude, $k_0$ denotes the initial
position of the spectral peak, $\alpha$ is the subinertial range slope
(here always $\alpha=4$), and the inertial range has a $k^{-5/3}$ spectrum.
Note that $\AAA$ is by construction periodic.
Therefore, $\BB=\nab\times\AAA$ has zero mean field.
At the end of this paper, we also briefly discuss a case with a finite
mean magnetic field.

The strength of the magnetic field can be characterized by the Alfv\'en
speed, $\vA=\Brms/\sqrt{\mu_0\rho_0}$, which is here based on the mean
density.
For $\varsigma\neq0$, we have a finite magnetic helicity and expect
then the decay to be governed by both the Hosking integral and the mean
magnetic helicity density.

In \Eq{SpB}, $\Sp(\cdot)$ denotes a shell integrated spectrum.
This operation will also be applied to the local, gauge-dependent
magnetic helicity density $h=\AAA\cdot\BB$, so that
$\Sp(h)=(k^2/8\pi^3L^3)\oint_{4\pi}|\tilde{h}|^2\,\dd\Omega_k$.
The tilde marks a quantity in Fourier space, and $\Omega_k$ is the
solid angle in Fourier space, so that $\int\Sp(h)\,\dd k=\bra{h^2}$,
and likewise for $\int\Sp(\BB)\,\dd k=\bra{\BB^2}$.
Owing to the integration over shells in three-dimensional wavenumber
space, the spectrum of a spatially random ($\delta$ correlated) field
is proportional to $k^2$.
This is indeed the case for a globally non-helical field, where
$\bra{h}=0$.

\subsection{Definitions of the Hosking integral}

The Hosking integral $I_{\rm H}$ is defined as the asymptotic limit of
the magnetic helicity density correlation integral,
\begin{equation}
{\cal I}_{\rm H}(R)=\int_{V_R}\bra{h(\xx)h(\xx+\rr)} \, \dd^3r,
\label{calI}
\end{equation}
for scales $R$ large compared with the correlation length $\xiM$ of the
turbulence, but small compared with the system size $L$.
Here, $V_R=4\pi R^3/3$ is the volume of a sphere of radius $R$.
For small values of $R$, the function ${\cal I}_{\rm H}(R)$ increases
proportional to $R^3$, but for large $R$, it levels off when there is
no net magnetic helicity.
However, as explained in \cite{HS21}, this is different for finite
magnetic helicity, as is discussed below.
In practice, the value of $R$ is chosen empirically and must always be
small compared with the size of the domain.

\cite{Zhou+22} devised and compared different methods for computing
$\mathcal{I}_\mathrm{H}(R)$.
These methods are all based on the Fourier transform of $h$.
Particularly simple is what they called the box-counting method for
a spherical volume with radius $R$.
This allowed them to rewrite \Eq{calI} as a weighted integral over $\Sp(h)$,
\begin{equation}
\mathcal{I}_\mathrm{H}(R)=\int_0^\infty w(k,R)\, \Sp(h) \, \text{d}k,
\label{eq:IHsph}
\end{equation}
where
\begin{equation}
w(k,R)=\frac{4\pi R^3}{3}
\left[\frac{6j_1(kR)}{kR}\right]^2,
\label{eq:wsph}
\end{equation}
and $j_1(x)=(\sin x-x\cos x)/x^2$ is a spherical Bessel function.

\subsection{Input and diagnostic parameters of the model}

Important input parameters of the model are the ratio of the initial Alfv\'en
speed to the sound speed, $\vAz/\cs$.
In the presence of an imposed mean field,
$\BB_\mathrm{m}=B_\mathrm{m}\xxx$, a case discussed at the end of the
paper, the corresponding Alfv\'en speed is denoted by $\vAm$.
To obtain information about the turbulent decay that is independent
of the size and shape of the computational domain, we must choose the
value of $k_0/k_1$ to be sufficiently large.
However, it should also not be chosen too large, because it would diminish
the range of wavenumbers between $k_0$ and the largest wavenumber in the
domain, which is called the Nyquist wavenumber, $k_\mathrm{Ny}=k_1 N/2$,
where $N$ is the number of meshpoints.
The sensitivity of the results on the choice of $k_0$ has been studied
on various occasions \citep[e.g.,][]{Zhou+22}.
A reasonable compromise that still allows for sufficiently large Reynolds
numbers seems to be $k_0/k_1=60$.
This is the value that is used for the main run in the present paper,
but we also present some results with $k_0/k_1=30$; see
\Tab{Runs} for a summary of runs presented in this paper.

\begin{table}\caption{
Summary of runs presented in this paper.
The arrows indicate the change from the beginning to the end of the run.
}\vspace{12pt}\centerline{\begin{tabular}{cccccccccc}
Run & $k_0/k_1$ & $\varsigma$ & $\vAz/\cs$ & $\vAm/\cs$ & $\Ma$ & $\Lu$ & $I_\mathrm{M}/2\xiM\EEM$ & $N^3$ & Figures \\
\hline
A & $60$ &   1   & 0.57 &  0  & $0.20\to0.03$  & $1000\to2500$ & $0.95\to0.94$ & $1152^3$ &   1     \\
B & $30$ & 0.003 & 0.21 &  0  & $0.10\to0.005$ & $2700\to1500$ & $0.01\to0.13$ & $1024^3$ & 2, 3, 8 \\
C & $60$ & 0.01  & 0.20 &  0  & $0.10\to0.001$ & $1300\to 800$ & $0.06\to0.78$ & $1024^3$ & 4--8    \\
D & $60$ &   0   & 0.22 & 0.1 & $0.11\to0.001$ & $ 800\to  70$ &        0      & $1024^3$ &   9     \\
\label{Runs}\end{tabular}}\end{table}

To characterize the degree of compressibility and the vigor of turbulence,
we quote the Mach and Lundquist numbers,
\begin{equation}
\Ma=\urms/\cs,\quad
\Lu=\vA/\eta k_0.
\end{equation}
Since our model is spatially homogeneous, it can be characterized by
the magnetic energy and helicity spectra, $\EM(k,t)$ and $\HM(k,t)$,
respectively.
They are normalized such that their integrals give
the mean magnetic energy and helicity densities,
$\EEM\equiv\int\EM(k,t)\,\dd k=\bra{\BB^2}/2\mu_0$ and
$I_\mathrm{M}\equiv\int\HM(k,t)\,\dd k=\bra{\AAA\cdot\BB}$,
respectively.\footnote{The name $I_\mathrm{M}$ has been chosen here
to mark its important role as an ideal invariant and to highlight its
usage analogously to that of the Hosking integral $I_\mathrm{H}$.}

The position of the peak of the spectrum is characterized by the inverse
magnetic integral scale, $k_\mathrm{peak}=\xiM^{-1}$, where $\xiM$
is here defined as
\begin{equation}
\xiM=\left.\int k^{-1}\,\EM(k,t)\,\dd k\,\right/\int \EM(k,t)\,\dd k.
\end{equation}
Of particular importance are the time dependencies $\xiM(t)$ and
$\EEM(t)$, which, in turn, are characterized by the instantaneous scaling
exponents $q(t)=\dd\ln\xiM/\dd\ln t$ and $p(t)=-\dd\ln\EEM/\dd\ln t$.
The relative magnetic helicity can be computed as the non-dimensional
ratio $I_\mathrm{M}/2\xiM\EEM$, which is between $-1$ and $+1$.

The relevant information that quantifies the Hosking integral is the
first nonvanishing coefficient in the Taylor expansion,
\begin{equation}
\left.\Sp(h)\right|_{k\to0}=\frac{I_\mathrm{H}}{2\pi^2}k^2+...\;;
\label{ExpSph}
\end{equation}
see \cite{HS21}, \cite{Scheko22}, and \cite{Zhou+22} for details.
This is also the primary method used here to determine the value
of $I_\mathrm{H}$; see \cite{Zhou+22} for a comparison between
different methods.
We confirm that $2\pi^2\Sp(h)/k^2$ has an approximately flat part
for small values of $k$ and use its value at $k=k_1$ to measure
$I_\mathrm{H}$.
Below, we also confirm that $I_\mathrm{H}$ is nearly independent of time;
see \cite{Zhou+22} for quantitative assessment of its invariance in the
ideal limit.
Note that, since $\int\Sp(h)\,\dd k=\bra{h^2}$, which has dimensions
$(\cm^3\s^{-2})^2$, $\Sp(h)$ has dimensions, $\cm^7\s^{-4}$ and therefore
$I_\mathrm{H}$ has dimensions $\cm^9\s^{-4}$, as expected.

To facilitate comparison with other work, it is useful to present our
results in nondimensional form.
The time used in the numerical simulations is made nondimensional by
plotting the evolution versus $\cs k_1 t$, which is convenient for
numerical reasons, because $\cs$ and $k_1$ are constant in time.
However, physically more meaningful would be a nondimensionalization
by using the Alfv\'en speed and the inverse correlation length.
Both are time dependent, but the values $v_\mathrm{Ae}$ and $k_\mathrm{e}$
at the end of the simulations seem to be most meaningful.

\subsection{Run time and scale separation}

To obtain meaningful results, two important constraints need to be obeyed.
First, the value of $\Lu$ needs to be large enough so that we are
in the regime of developed turbulence.
Second, the subinertial range must always be large enough so that, by the
end of the run, its slope is not affected by finite size effects of the
computational domain.
This automatically limits the maximum run time below which our results
can still be meaningful.
Both constraints can only be obeyed in the limit of infinite resolution.
In practice, the largest resolution that is presently feasible is
typically $2048^3$ meshpoints \citep{Zhou+22}, but this large resolution
does already constrain the number of experiments that can reasonably
be performed.
Therefore, we use for most of our simulations a lower resolution of
$N^3=1024^3$ meshpoints.
In that case, the largest wavenumber in the domain is 
$k_\mathrm{Ny}=k_1 N/2=512\,k_1$.
As discussed above, this led us to the compromise of choosing
the values 30 and 60 for the scale separation ratio $k_0/k_1$, so
$k_0/k_\mathrm{Ny}=17$--$8.5$, leaving barely enough dynamical range
for turbulence to develop.

\section{Results}

In the present context, we have to deal with two conserved quantities,
namely the Hosking integral $I_\mathrm{H}$ and the mean magnetic helicity
density $I_\mathrm{M}=\bra{\AAA\cdot\BB}$.
The former case has been studied extensively in recent years.
Specifically, \cite{BL23} and \cite{BSV23} found
\begin{equation}
\xiM(t)\approx0.12\,I_{\rm H}^{1/9}\,t^{4/9},\quad
\EEM(t)\approx3.7\,I_{\rm H}^{2/9}\,t^{-10/9},\quad
\EM(k,t)\la0.025\,I_{\rm H}^{1/2}\,(k/k_0)^{3/2}.
\label{HoskingFits}
\end{equation}
The hope is that the coefficients in these expressions are universal, but
it should be noted that they have not yet been verified in other contexts.

\subsection{Decay controlled by mean and fluctuating magnetic helicities}

In the helical case with $I_\mathrm{M}\neq0$, we have $\xiM\propto
t^{2/3}$ and $\EEM\propto t^{-2/3}$ \citep{Hatori84, BM99, BK17}.
In the present context, the pre-factors are important.
Using the data from figures~1(c) and 2(c) of \cite{BK17}, here
referred to as Run~A, we find
\begin{equation}
\xiM(t)\approx0.12\,I_{\rm M}^{1/3}\,t^{2/3},\quad
\EEM(t)\approx4.3\,I_{\rm M}^{2/3}\,t^{-2/3},\quad
\EM(k,t)\la0.7\,I_{\rm M}.
\label{HelicityFits}
\end{equation}
Generally, we can write
\begin{equation}
\xiM(t)=C_{i}^{(\xi)}\,I_{i}^{\sigma}t^q,\quad
\EEM(t)=C_{i}^{({\cal E})}\,I_{i}^{2\sigma}t^{-p},\quad
\EM(k)=C_{i}^{(E)}\,I_{i}^{(3+\beta)\sigma}(k/k_0)^{\beta},
\label{GeneralFits}
\end{equation}
where the index $i$ in the integrals $I_i$ and the coefficients
$C_{i}^{(\xi)}$, $C_{i}^{({\cal E})}$, and $C_{i}^{({E})}$ stands for M or
H for magnetic helicity and Hosking scalings, respectively, and $\sigma$
is the exponent with which length enters in $I_{i}$: $\sigma=1/3$ for
the magnetic helicity density ($i={\rm M}$) and $\sigma=1/9$ for the
Hosking integral ($i={\rm H}$); see \Tab{TExpSummary} for a summary of
the coefficients.\footnote{In equation~(3.3\textit{c}), we have here corrected
a typo in equation~(12\textit{c}) of \cite{BL23} and equation~(3.1\textit{c}) of
\cite{BSV23}, where the exponent on $I_{i}$ was incorrectly stated as
$(3+\beta)/\sigma$ instead of $(3+\beta)\sigma$, but the calculations
were done correctly.}

\begin{table}\caption{
Summary of the coefficients characterizing the decays governed
by the conservation of magnetic helicity ($i=\mathrm{M}$) and
the Hosking integral ($i=\mathrm{H}$).
}\vspace{12pt}\centerline{\begin{tabular}{cccccccc}
$i$ & $\beta$ & $q$ & $p$ & $\sigma$ & $C_{i}^{(\xi)}$ & $C_{i}^{({\cal E})}$ & $C_{i}^{(E)}$ \\
\hline
M &  0  & 2/3 &  2/3 & 1/3 & 0.13 & 4.1 & 0.7   \\
H & 3/2 & 4/9 & 10/9 & 1/9 & 0.12 & 3.7 & 0.025 \\
\label{TExpSummary}\end{tabular}}\end{table}

\begin{figure}\begin{center}
\includegraphics[width=\columnwidth]{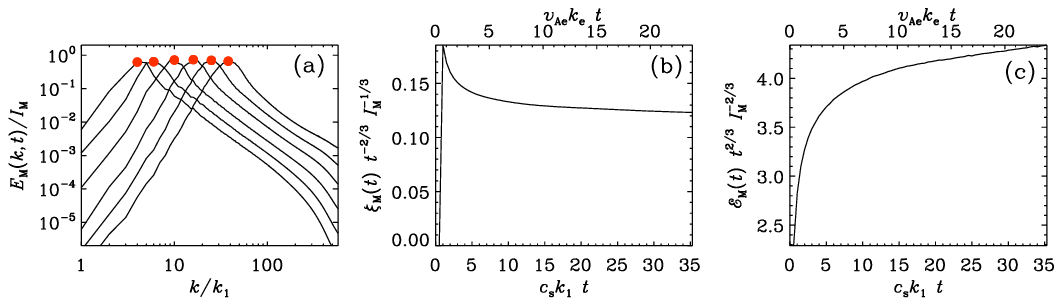}
\end{center}\caption[]{
(a) Magnetic energy spectra, as well as compensated evolutions
of (b) $\xiM(t)$ and (c) $\EEM(t)$ for the maximally helical run
of figure~2(c) of \cite{BK17}, here referred to as Run~A.
In (a), the red symbols denote the spectral peaks.
}\label{pxi}\end{figure}

In \Fig{pxi}, we show the magnetic energy spectra, as well as compensated
evolutions of $\xiM(t)$ and $\EEM(t)$ for the maximally helical run (Run~A).
We see that the peak of $\EEM(t)$ remains underneath a nearly flat
envelope (its slope is $\beta=0$), as is expected for a fully helical
turbulent decay at late times.
The compensated evolutions of $\xiM(t)$ and $\EEM(t)$
are not yet fully converged toward the end of that run (the
lines are not yet flat).
This is partially caused by the insufficient scale separation between the
box wavenumber $k_1$ and that of the spectral peak by the end of the run.
Nevertheless, we can read off the approximate
values $C_\mathrm{M}^{(\xi)}\approx0.12$ and
$C_\mathrm{M}^{({\cal E})}\approx4.3$ towards the end of the run.
The values of these coefficients are revisited later in this paper.

\begin{figure}\begin{center}
\includegraphics[width=\columnwidth]{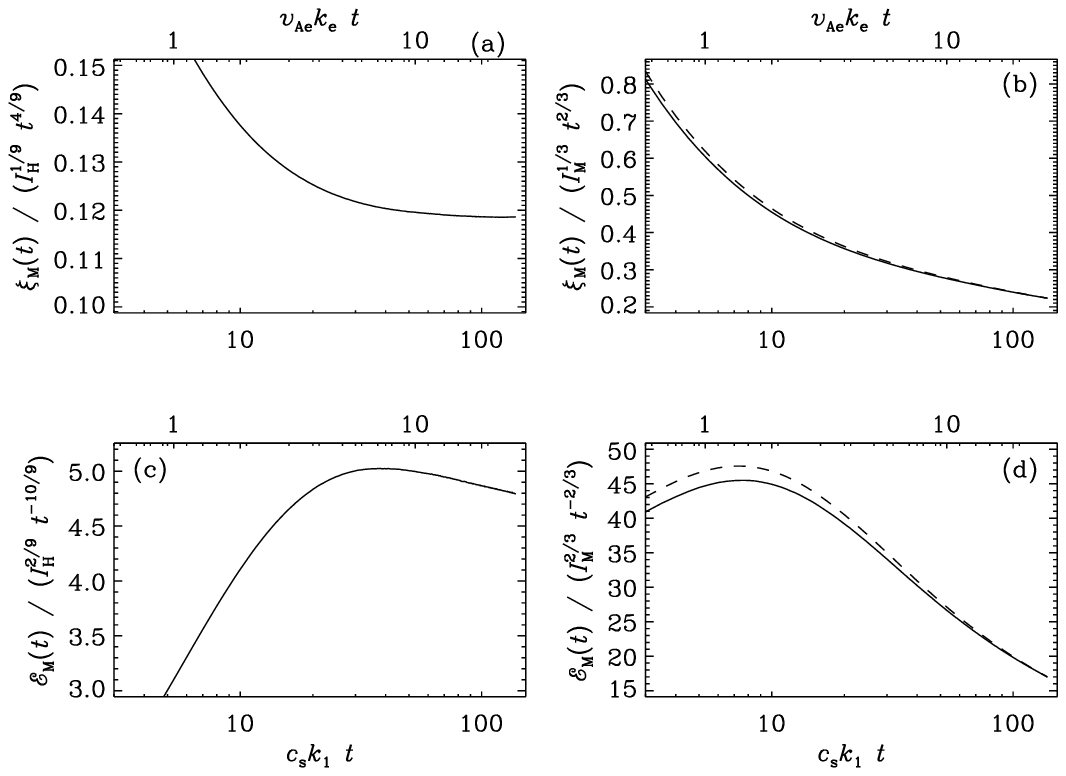}
\end{center}\caption[]{
Evolution of $\xiM(t)$ (upper row) and $\EEM(t)$ (lower row) for Run~B
with $k_0/k_1=30$ and $\varsigma=0.003$, compensated by the expected
evolution if the decay is controlled either by $I_\mathrm{H}$ (left
column) or by $I_\mathrm{M}$ (right column).
The dashed line denotes the use of $I_\mathrm{M}$ at the end of the run,
while for the solid line, the time-dependent value was taken.
}\label{ppxi_hosk_M1024c_sig0p003_k30}\end{figure}

In \Fig{ppxi_hosk_M1024c_sig0p003_k30}, we again show the compensated
evolutions of $\xiM(t)$ and $\EEM(t)$, but now for Run~B, which is nearly
perfectly nonhelical ($\varsigma=0.003$) and has $k_0/k_1=30$.
The resulting coefficients are close to those estimated
previously, namely $C_\mathrm{H}^{(\xi)}\approx0.12$ and
$C_\mathrm{H}^{({\cal E})}\approx4.7$.
This supports the previous hypotheses of \cite{BL23} and \cite{BSV23}
that these coefficients may indeed be universal.

\begin{figure}\begin{center}
\includegraphics[width=\columnwidth]{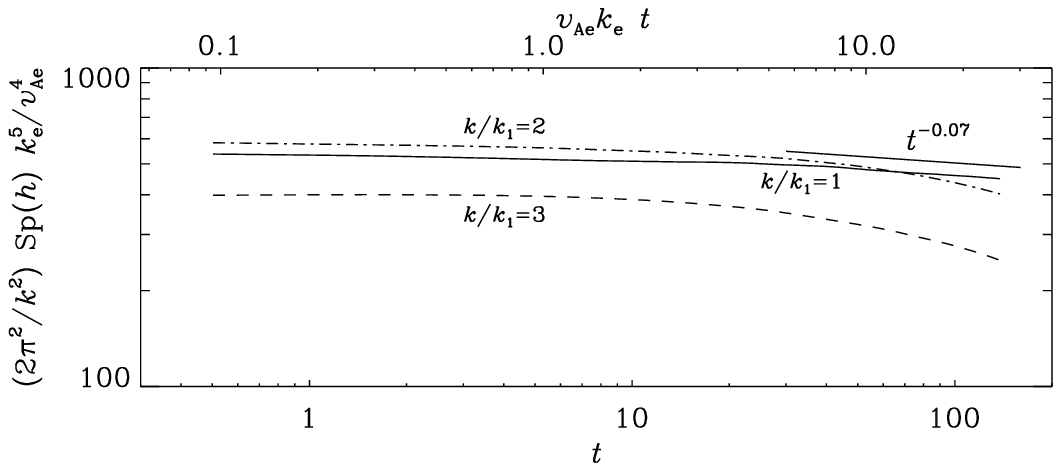}
\end{center}\caption[]{
Evolutions of $(2\pi^2/k^2)\,\Sp(h)$, normalized by
$v_\mathrm{Ae}^4/k_\mathrm{e}^5$, for $k/k_1=1$ (solid line),
2 (dashed-dotted line), and 3 (dashed line), for the nearly nonhelical
Run~B with $\varsigma=0.003$ and $k_0/k_1=30$.
}\label{calc_hosk_M1024c_sig0p003_k30}\end{figure}

To determine the value of $I_\mathrm{H}(t)$, we plot
in \Fig{calc_hosk_M1024c_sig0p003_k30} the evolutions of
$(2\pi^2/k^2)\,\Sp(h)$ (normalized by $v_\mathrm{Ae}^4/k_\mathrm{e}^5$) for
$k/k_1=1$, 2, and 3 for Run~B.
We see that for $k/k_1=1$, the result shows a nearly negligible decline
proportional to $t^{-0.07}$.
Note that, in units of $v_\mathrm{Ae}^4/k_\mathrm{e}^5$, the value of
$I_\mathrm{H}$ is about 500.

\subsection{Decay controlled by $I_\mathrm{M}$ and $I_\mathrm{H}$}

If both $I_\mathrm{M}$ and $I_\mathrm{H}$ control the decay, we
have a combination of the two decay laws such that the late times
are always controlled by the more strongly conserved quantity,
i.e., by $I_\mathrm{M}$.
One might expect that the resulting expression for the combination of
the decay laws \eq{HoskingFits} and \eq{HelicityFits} is given by the
sum of both expressions.
This would be analogous to the way how in radiation transport the
cooling time is given by the sum of the cooling times for the
optically thick and thin cases; see equation~(7) of \cite{BD20}.
In the present case, this would translate to
\begin{equation}
\xiM\approx 0.12 \, I_\mathrm{M}^{1/3} \, t^{2/3}
+0.12 \, I_\mathrm{H}^{1/9} \, t^{4/9},
\label{combined1xiM}
\end{equation}
\begin{equation}
\EEM\approx 4.3 \, I_\mathrm{M}^{2/3} \, t^{-2/3}
+3.7 \, I_\mathrm{H}^{2/9} \, t^{-10/9}.
\label{combined1EEM}
\end{equation}
Since the second terms involving $I_\mathrm{H}$ are initially larger,
but their contributions to $\xiM$ grow more slowly
and that to $\EEM$ decay faster than the first terms,
one expects their contributions to become subdominant after some time.
Thus, the magnetic helicity will always survive and be the dominant
contribution to explaining the decay.

\begin{figure}\begin{center}
\includegraphics[width=\columnwidth]{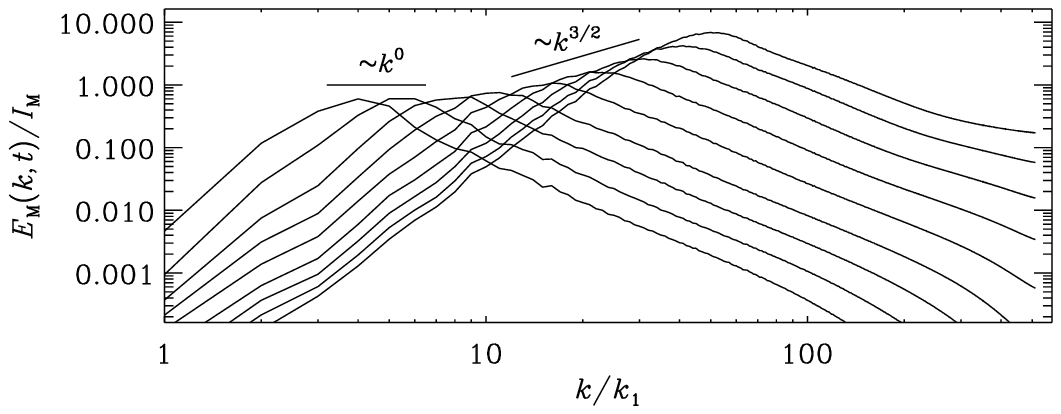}
\end{center}\caption[]{
Magnetic energy spectra for Run~C with $k_0/k_1=60$ and $\varsigma=0.01$
at times $v_\mathrm{Ae} k_\mathrm{e}\,t=0.07$, 0.18, 0.40, 0.82, 1.65,
3.3, 6.1, 11.1, and 20.7.
}\label{pspec_M1024c_sig0p001_k60}\end{figure}

To examine now a run where the decay is controlled by both $I_\mathrm{M}$
and $I_\mathrm{H}$, we now increase the initial fractional helicity
slightly from 0.003 to 0.01; see \Fig{pspec_M1024c_sig0p001_k60} for
magnetic energy spectra at different times for Run~C.
Note that the peaks of the spectra evolve at first underneath an
envelope with the slope $\beta=3/2$, as expected for a decay controlled
by $I_\mathrm{H}$.
At later times, however, the envelope becomes flat (slope $\beta=0$), 
as expected for a decay controlled by $I_\mathrm{M}$.

\begin{figure}\begin{center}
\includegraphics[width=\columnwidth]{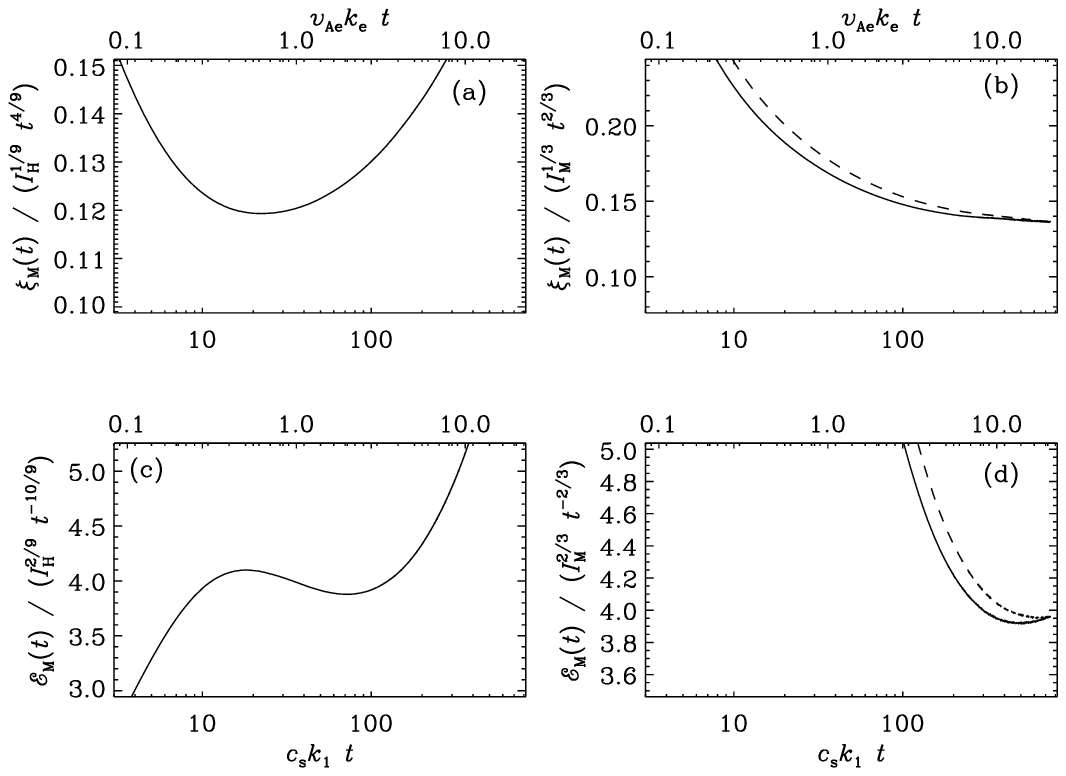}
\end{center}\caption[]{
Similar to \Fig{ppxi_hosk_M1024c_sig0p003_k30},
but for Run~C with $k_0/k_1=60$ and $\varsigma=0.01$.
}\label{ppxi_M1024c_sig0p001_k60}\end{figure}

In \Fig{ppxi_M1024c_sig0p001_k60}, we show the evolutions of $\xiM(t)$
and $\EEM(t)$ for Run~C, compensated by $t^{-2/3}$ and $t^{2/3}$,
respectively, as well as $t^{-4/9}$ and $t^{10/9}$, respectively.
We see that now the curves compensated by $t^{-2/3}$ and $t^{2/3}$,
respectively, become nearly constant, as expected for a decay that is
governed by magnetic helicity conservation.
Specifically, we find
$\xiM/(I_{\rm M}^{1/3}\,t^{2/3})\approx0.14$ and
$\EEM/(I_{\rm M}^{2/3}\,t^{-2/3})\approx4.0$.
During a short intermediate interval, however, we see that the
curves compensated by $t^{-4/9}$ and $t^{10/9}$, respectively,
show brief plateaus around $v_\mathrm{Ae}k_\mathrm{e}t=1$ with
$\xiM/(I_{\rm H}^{1/9}\,t^{4/9})\approx0.12$ and
$\EEM/(I_{\rm H}^{2/9}\,t^{-10/9})\approx4.0$.

\begin{figure}\begin{center}
\includegraphics[width=\columnwidth]{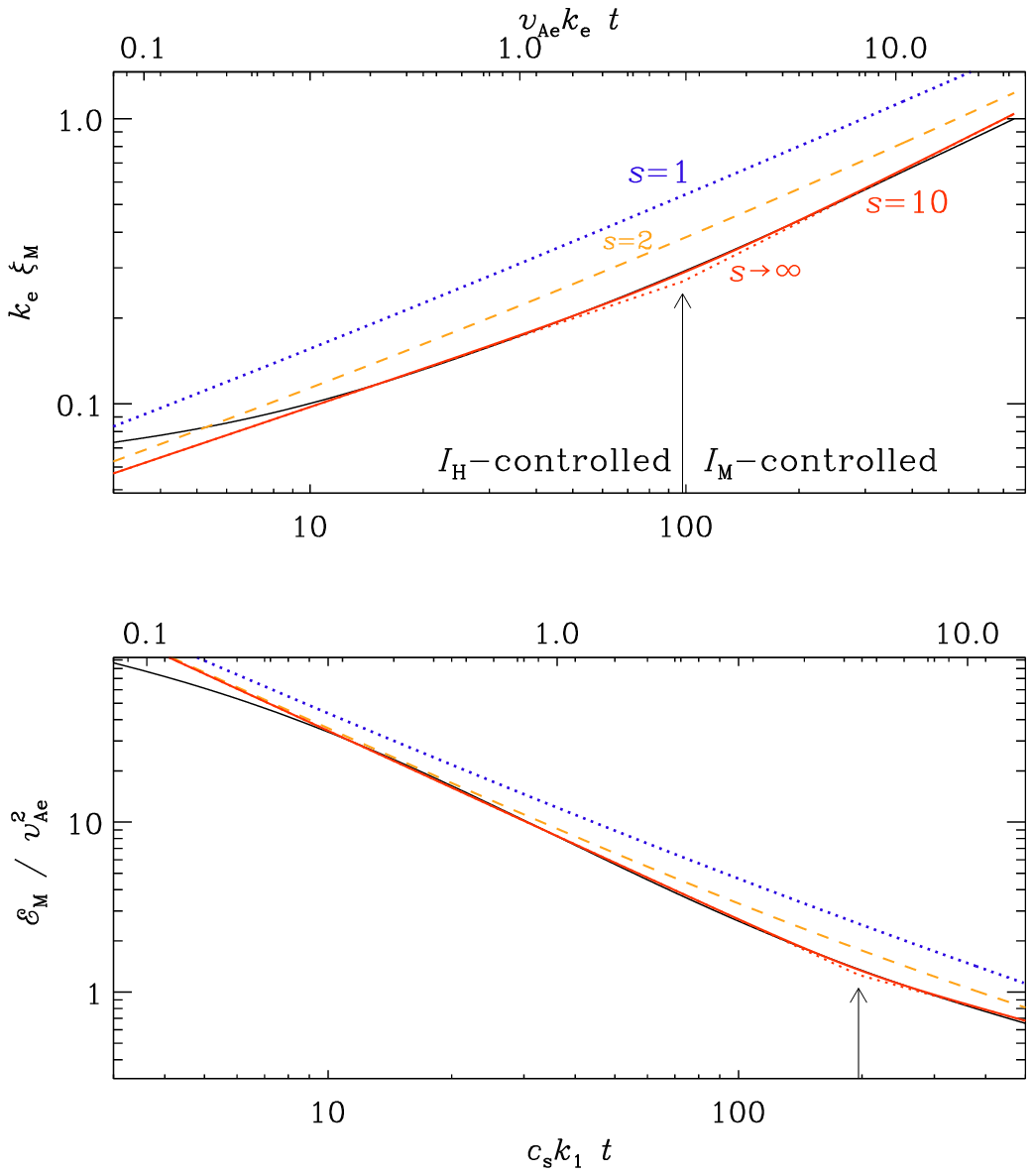}
\end{center}\caption[]{
Decay of magnetic energy (black line) and the fit given by
\Eq{combined1EEM} (dotted blue line, denoted by $s=1$) as well as
\Eq{combined2EEM} with $s=2$ (dashed orange line) and $s=10$ (solid
red line).
The dotted red line corresponds to the limit $s\to\infty$, as
realized by \Eqs{combined3xiM}{combined3EEM}.
}\label{pfit}\end{figure}

\subsection{Improved fits with $I_\mathrm{M}$ and $I_\mathrm{H}$}
\label{ImprovedFits}

We have seen that the limiting cases where the decay is controlled
either by $I_\mathrm{M}$ or by $I_\mathrm{H}$ are well reproduced by
\Eq{GeneralFits}.
It turns out, however, that the combined fits given by
\Eqs{combined1xiM}{combined1EEM} are not very accurate.
Improved fits can be obtained by using large weighting exponents for
both contributions, i.e.,
\begin{equation}
\xiM\approx \left[
\left(0.12 \, I_\mathrm{M}^{1/3} \, t^{2/3}\right)^s +
\left(0.14 \, I_\mathrm{H}^{1/9} \, t^{4/9}\right)^s\right]^{1/s},
\label{combined2xiM}
\end{equation}
\begin{equation}
\EEM\approx \left[
\left(4.0 \, I_\mathrm{M}^{2/3} \, t^{-2/3}\right)^s +
\left(4.0 \, I_\mathrm{H}^{2/9} \, t^{-10/9}\right)^s\right]^{1/s}.
\label{combined2EEM}
\end{equation}
The result is shown in \Fig{pfit} for Run~C, where we show that $s=10$
yields satisfactory fits, while $s=2$ and $s=1$ (our original hypothesis)
are poor.
The fact that the coefficients for both parts are different from those
of the individual fits and that they happen to be $4.0$ in 
\Eq{combined2EEM}, but different from each other in \Eq{combined2xiM}
is probably just by chance and reflect that degree of uncertainty of
these values.

It is important to emphasize that the limit $s\to\infty$ corresponds to
\begin{equation}
\xiM\approx \max\left(
0.12 \, I_\mathrm{M}^{1/3} \, t^{2/3},\;
0.14 \, I_\mathrm{H}^{1/9} \, t^{4/9}\right),
\label{combined3xiM}
\end{equation}
\begin{equation}
\EEM\approx \max\left(
4.0 \, I_\mathrm{M}^{2/3} \, t^{-2/3},\;
4.0 \, I_\mathrm{H}^{2/9} \, t^{-10/9}\right).
\label{combined3EEM}
\end{equation}
These expressions yield discontinuities in the derivative.
An advantage of such expressions is that one can clearly see the regimes
of validity of both expressions.
The critical times when characterizing the cross-over from Hosking
scaling to magnetic helicity scaling are given by
\begin{equation}
t_\xi\approx(0.12/0.14)^{9/2}\left(I_\mathrm{H}/I_\mathrm{M}^3\right)^{1/2}
\approx0.50\,I_\mathrm{H}^{1/2}I_\mathrm{M}^{-3/2},
\label{combined4xiM}
\end{equation}
\begin{equation}
t_{\cal E}\approx
I_\mathrm{H}^{1/2}I_\mathrm{M}^{-3/2}.
\label{combined4EEM}
\end{equation}
It would be plausible to assume that both times should equal each other.
The fact that they are not equal to each other might hint, again,
at the possibility that the precise values of these coefficients are
still uncertain.
On the other hand, looking at \Fig{ppxi_M1024c_sig0p001_k60}, it is
actually true that $\xiM$ approaches the $I_\mathrm{M}$-dominated scaling
by a factor two earlier than $\EEM$.
A possible explanation for this behavior could lie in the fact that the
spectral shapes change as the system becomes fully helical.
We return to this in \Sec{Collapsed}, where we discuss the spectral
shapes in more detail.

We recall that an essential assumption in our dimensional argument was
the fact that the magnetic field is understood to be in Alfv\'en units
and thus has dimensions of $\cm\s^{-1}$.
In neutron star crusts, by contrast, where the magnetic field is governed
by the Hall effect \citep{GR92}, it has units of $\cm^2\s^{-1}$, so
$[I_\mathrm{M}]=\cm^5\s^{-2}$ and $[I_\mathrm{H}]=\cm^{13}\s^{-4}$
\citep{Bra23}, so $t_\xi$ and $t_{\cal E}$ are now proportional to
$I_\mathrm{H}^{5/6}I_\mathrm{M}^{-13/6}$, so both exponents are larger
than in the ordinary magnetohydrodynamic case; cf.\ \Eqs{combined4xiM}{combined4EEM}.
An example of the corresponding switch between the two regimes was
presented in figure~10(b) of \cite{Bra20}.

\subsection{Collapsed spectra}
\label{Collapsed}

\begin{figure}\begin{center}
\includegraphics[width=\columnwidth]{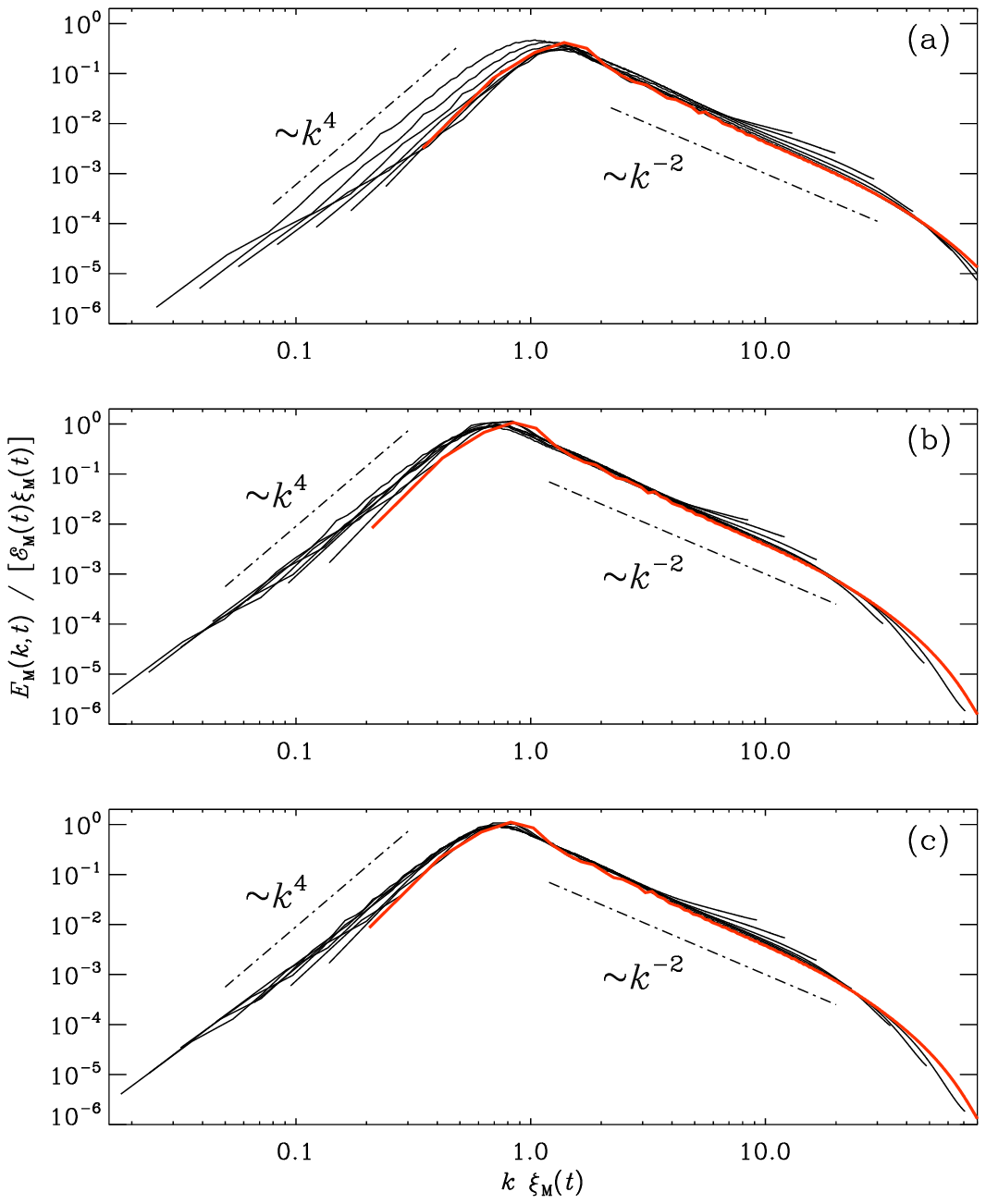}
\end{center}\caption[]{
Magnetic energy spectra similar to \Fig{pspec_M1024c_sig0p001_k60},
but the abscissa is scaled with $\xiM(t)$ and the ordinate with
$[\EEM(t)\xiM(t)]^{-1}$, where \Eqs{combined2xiM}{combined2EEM} are used
with $s=1$ in panel (a), and with $s\to\infty$ in panel (b).
In panel~(c), the actual values of $\xiM(t)$ and $\EEM(t)$ are used.
The last time is shown as a thick red line.
}\label{pspec_M1024c_collapse_sig0p001_k60}\end{figure}

The quality of the fits of \Eqs{combined2xiM}{combined2EEM} for $s=1$
and $s\to\infty$ can be examined further by computing compensated spectra.
This is shown in \Fig{pspec_M1024c_collapse_sig0p001_k60},
where the abscissa is scaled with $\xiM(t)$ and the ordinate with
$[\EEM(t)\xiM(t)]^{-1}$.
We see that the collapse in \Fig{pspec_M1024c_collapse_sig0p001_k60}(b),
where $s\to\infty$, is much better than that in
\Fig{pspec_M1024c_collapse_sig0p001_k60}(a), where $s=1$, and it is
almost as good as that in \Fig{pspec_M1024c_collapse_sig0p001_k60}(c),
where the actual values of $\xiM(t)$ and $\EEM(t)$ are used.
This supports our finding that in the fractionally helical case, the
magnetic energy and correlation length are approximately given by the
maximum of the values for the purely helical and purely nonhelical cases,
and not by their sum, as might naively have been expected.

As alluded to in \Sec{ImprovedFits}, there is a change in the
shape of the spectrum as the system becomes fully helical.
In particular, the position of the peak appears for
slightly larger values of $k\xiM(t)$ at later times; see
\Fig{pspec_M1024c_collapse_sig0p001_k60}(c).
Thus, the value of $\xiM(t)$ is slightly overestimated, which would
explain the smaller value of $t_\xi$ compared with $t_{\cal E}$.

\subsection{Comparison with earlier work}
\label{EarlierWork}

As mentioned in the introduction, the switchover time from nonhelically
to helically dominated decay has been studied by \cite{Tevzadze+12} under
the assumption that $p=1$ and $q=1/2$ \citep{CHB01} instead of $p=10/9$
and $q=4/9$, as now motivated by the conservation of the Hosking integral.
The basic idea is to assume that at the time of switchover, $t_\ast$,
the real-space realizability condition \citep{Biskamp+03, Kahniashvili+10}
is saturated, i.e., $2\xiM(t_\ast)\EEM(t_\ast)/\rho_0=I_\mathrm{M}$.
Next, inserting $\xiM(t_\ast)=\xiM(t_0)\,(t_\ast/t_0)^q$ and
$\EEM(t_\ast)=\EEM(t_0)\,(t_\ast/t_0)^{-p}$, we find
$2\xiM(t_0)\EEM(t_0)/\rho_0\,(t_\ast/t_0)^{-(p-q)}=I_\mathrm{M}$,
and therefore
\begin{equation}
t_\ast=t_0\,\left[2\xiM(t_0)\EEM(t_0)/I_\mathrm{M}\rho_0\right]^{1/(p-q)}.
\label{tast_T+12}
\end{equation}
For $p=1$ and $q=1/2$, we have $1/(p-q)=2$ and 
recover the result of \cite{Tevzadze+12},
while for $p=10/9$ and $q=4/9$, we have $1/(p-q)=3/2$.
Comparing with \Eq{combined4EEM}, we see that $I_\mathrm{M}$ enters with
the same exponent $3/2$, and that the remainder can be identified with
\begin{equation}
I_\mathrm{H}=\left[2\xiM(t_0)\EEM(t_0)\right]^3 t_0^2.
\label{t0}
\end{equation}
This suggests that $I_\mathrm{H}$ is related to $\xiM$ and $\EEM$,
but the problem is that $t_0$ is not straightforwardly related to the
Alfv\'en time $\xiM/\vA$, where $\vA^2=2\EEM/\rho_0$.
Indeed, \cite{Bra+24} found that there is a pre-factor $C_\mathrm{M}$
that increases with increasing magnetic Reynolds number.
Such a factor has been motivated based on magnetic reconnection arguments
\citep{HS23}.
Thus, inserting $t=C_\mathrm{M}\xiM/\vA$, we find
\begin{equation}
I_\mathrm{H}=C_\mathrm{M}^2\xiM^5\vA^4.
\label{tb}
\end{equation}
The facts that $C_\mathrm{M}$ enters quadratically and approaches values
in the range 20--50 for large magnetic Reynolds numbers explains why
$I_\mathrm{H}$ strongly exceeds the naive estimate $\xiM^5\vA^4$.
Interestingly, \cite{Zhou+22} found that part of the large excess over
the naive estimate is related to non-Gaussianity.
Another smaller part has to do with the spectral shape.
Linking the value of $C_\mathrm{M}$ to non-Gaussianity of the
magnetic field provides a new clue to the question of why there is
a resistivity-dependent relation between decay and Alfv\'en times in
hydromagnetic turbulence.

\begin{figure}\begin{center}
\includegraphics[width=\columnwidth]{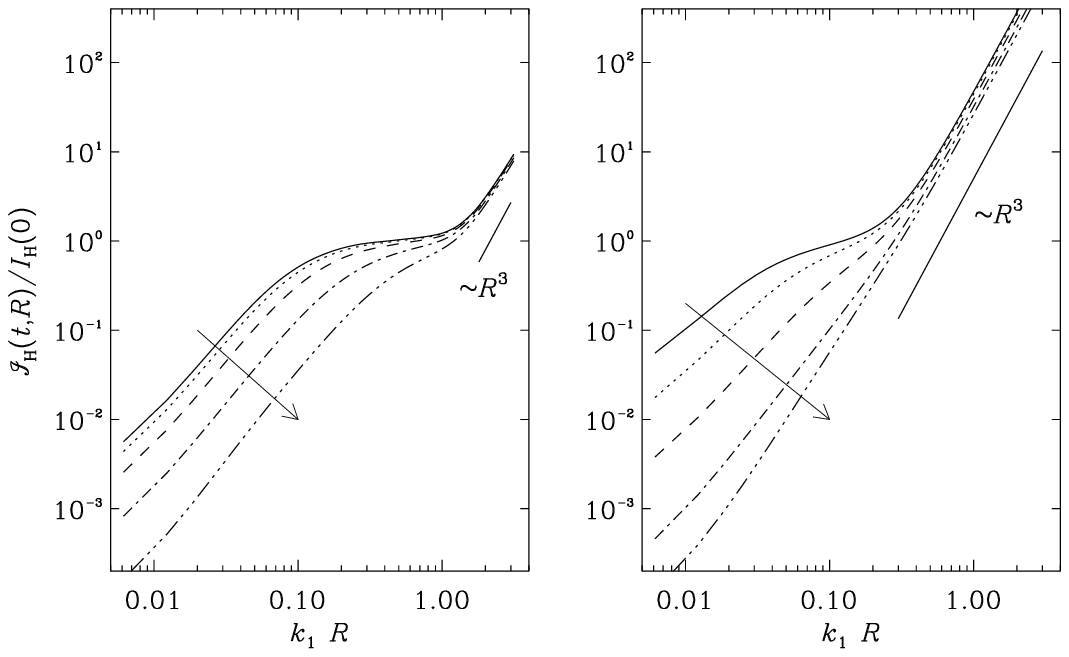}
\end{center}\caption[]{
Box-counting result for $\mathcal{I}_\mathrm{H}(R)$ for runs~B and C in
the left- and right-hand panels.
Note the plateau for intermediate values of $R$ at early times.
}\label{psaff_comp}\end{figure}

\subsection{Can the switchover time be resistively limited?}

We know that the decay time, $\tau(t)=(-\dd\ln\EEM/\dd t)^{-1}=t/p(t)$
can be regarded as resistively limited when relating it to the Alfv\'en
time, $\tau_\mathrm{A}=\xiM/\vA$.
In particular, as alluded to in \Sec{EarlierWork}, it turns out that
$\tau/\tau_\mathrm{A}=C_\mathrm{M}(\Lu)$, which is a monotonically
increasing function of $\Lu$ that saturates near $\Lu_\ast$ at
$C_\mathrm{M}^\ast\approx50$ \citep{Bra+24}.
Such a relation was theoretically expected and has been associated with
magnetic reconnection \citep{HS23}.
However $C_\mathrm{M}(\Lu)$ was found to be independent of the value
of the magnetic Prandtl number, which raised doubts about this
interpretation.

The question now is whether the switchover time might also depend on
the value of $\Lu$.
Differentiating \Eq{combined2EEM} or \Eq{combined3EEM}, we see that
the decay time depends on whether $t<t_\ast$ or $t>t_\ast$ and is equal
to $9t/10$ or $3t/2$, respectively, but the switchover time itself is
unaffected by resistivity effects.
In other words, the decay time is always $t/p$, where the value of $p$
depends on the time.

There is also the possibility that for $\Lu<\Lu_\ast$, the exponent $p$
might depend on the order of the diffusion operator, i.e., on whether it
is proportional to $\nabla^2$ or some higher power; see \cite{Zhou+22}
for details.
However, this consideration only applies to the regime of low enough
values of $\Lu$ and it would still not affect the actual value of
$t_\ast$.

\subsection{Hosking scaling at intermediate length scales}

\cite{HS21} presented arguments that for finite magnetic helicity,
the Hosking scaling should only be obeyed at intermediate length scales.
To check this, we now use the box-counting method as described by
\Eq{eq:IHsph} to plot $\mathcal{I}_\mathrm{H}(t,R)$ at different times $t$.
The result is shown in \Fig{psaff_comp} and resembles the sketch provided
in figure~10 of \cite{HS21}.
We do indeed see a short plateau where the Hosking scaling can be
discerned for intermediate times.
Furthermore, at later times, we see the expected $R^3$ scaling over the
whole range of $R$.

\begin{figure}\begin{center}
\includegraphics[width=\columnwidth]{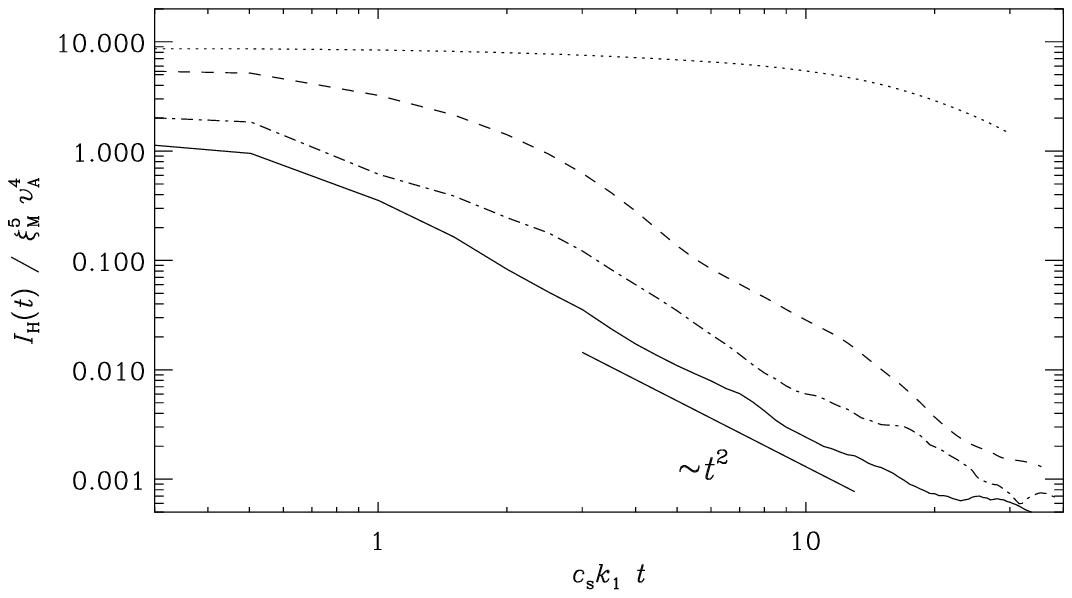}
\end{center}\caption[]{
Time dependence of $\Sp(h)$ for run~D (solid curve, $\vAm/\cs=0.1$)
and several cases with weaker mean field ($\vAm/\cs=0.05$ for the
dashed-dotted line, 0.02 for the dashed line, and 0.01 for the dotted
line).
}\label{ppcalc_hosk}\end{figure}

\subsection{Comment on the case of a finite mean field}

Our simulations presented above had zero mean field.
It is well known that in the presence of a mean field across
the entire domain, the magnetic helicity is no longer conserved
\citep{Berger97, BM04}; see \cite{Bran+20} for corresponding
decay simulations in the presence of a mean field.
To check whether the Hosking integral could still be meaningful in
such a case, we now present the time dependence of $I_\mathrm{H}(t)$
for run~D with different magnetic field strength, where a mean field
$\BB_\mathrm{m}=B_\mathrm{m}\xxx$ is now imposed, so the magnetic field
is given by $\BB=\BB_\mathrm{m}+\nab\times\AAA$.
As before, we evaluate $I_\mathrm{H}(t)=2\pi^2\Sp(h)/k^2$ at $k=k_1$.
The result is shown in \Fig{ppcalc_hosk} for four values of $\vAm$.
We see that $I_\mathrm{H}(t)$ is now decaying $\propto t^{-2}$, i.e., the Hosking
integral is not conserved.
Thus, with periodic boundary conditions, not only is $I_\mathrm{M}$
not conserved, but $I_\mathrm{H}$ is also not conserved.

\subsection{Comment on the case with chiral fermions}

As we have mentioned in the introduction, the Hosking integral
also describes the decay of helical turbulence in the presence of
chiral fermions if their chirality exactly balances the magnetic
helicity.
One may therefore ask whether a switchover to helical scaling could
also occur in this case of the initial balance being not perfect.
In \cite{Bra+23}, two cases of imbalanced chirality were already
presented.
In the case where the magnetic helicity exceeds the negative contribution
to the chirality, a helical decay scaling $\propto t^{-2/3}$ of the
magnetic energy was found,
thus supporting our expectation.
In the opposite case of an excess of fermion chirality, the time
evolution of the magnetic energy was more complicated and no
clear scaling suggestive of a helical decay was found.

\section{Conclusions}

The present work has shown that the decay laws for the combined case of
two conserved quantities is best represented not simply by the sum of
the individual laws, but that a good description of the numerical results
is obtained by taking the maximum between the two individual decay laws.
The switchover from one to the other decay law occurs earlier for
$\xiM(t)$ than for $\EEM(t)$.
This behavior is surprising, but confirmed by direct inspection of the
two time traces in \Fig{ppxi_M1024c_sig0p001_k60}(b) and (d) and perhaps
explained by changes in the shape of the magnetic energy spectrum during
an otherwise almost perfectly self-similar decay.

Comparing with earlier work on the switchover from one to the other
regime suggests that the ratio of the decay time to the Alfv\'en time
enters in such a relation.
This is remarkable, because in hydromagnetic turbulence the decay time
is known to be longer than the Alfv\'en time by a resistivity-dependent
factor of up to 50 \citep{Bra+24}.
This large factor might also explain why the value of the Hosking integral
is always found to strongly exceed the naive estimate $\xiM^5\vA^4$.
In other words, the reason why this simple formula underestimates
the value of the Hosking integral might be the occurrence of the same
resistivity-dependent factor that also occurs in the expression for the
Alfv\'en time.
However, as shown in \cite{Zhou+22}, also other factors enter that
involve the spectral shape.
It would therefore be interesting to revisit this question.

As alluded to in the introduction, an obvious astrophysical application
of our work is the decay of an initially bihelical magnetic field.
Such situations are important in proto-neutron stars after the
neutrino-driven convection ceases.
Although this was actually our initial motivation, we have not analyzed
this case any further, because the most important aspect turned out to
be the fact that the magnetic field has always fractional helicity in
such cases, which we have now addressed in the present paper.
A problem with the application to proto-neutron stars is of course the
fact that in stars, the magnetic field is inhomogeneous and the decay
is initially not yet magnetically dominated; see \cite{Bran+19} and
\cite{Uchida+24}.
Another obvious application is to the decay of primordial magnetic fields
during the radiation-dominated era of the early universe, which led to
the aforementioned work by \cite{Tevzadze+12}.

A more general question is that of a decay governed by two decay laws
and whether there are other useful examples where the physics discussed
in the present paper can be studied.
As far as turbulence is concerned, one might think of the Saffman
and Loitsyansky integrals, which represent the coefficients of the $k^2$
and $k^4$ terms in the Taylor expansion of the kinetic energy spectrum
\citep[see, e.g.,][]{Davidson00}.
An initial $k^4$ spectrum (for a vanishing Saffman integral)
might survive for some time, but neither of the two integrals is well
conserved, and the Saffman integral might become important at later
times when long-range interactions have occurred \citep{HS23b}.
This idea could also be applied to the magnetic case.

The case with an imposed magnetic field in a triply periodic domain is
known to be peculiar.
The mean magnetic helicity density of the remaining magnetic field
(without the imposed one) is not conserved \citep{Berger97}.
Although one can construct an additional quantity that takes the imposed
field into account \citep{Stribling+94}, it turns out that it is not
gauge-variant \citep{BM04}.
Our present work has shown that with an imposed mean magnetic field,
also the Hosking integral is no longer conserved and tends to zero.

In summary, the present work has extended our knowledge about the
Hosking integral, a remarkably useful quantity whose influence on many
aspects of decaying hydromagnetic turbulence can be understood based on
dimensional analysis.
Numerical simulations are used to pinpoint the values of the coefficients.
For several different systems, the set of these coefficients has been
found to be similar, suggesting that their values might be fundamental
quantities.
But more work is required to establish this more firmly.

\section*{Acknowledgements}

We thank Ivan Khaymovich and Geoff Vasil for inspiring discussions and
useful comments on our work.
We are also grateful to the three referees for their constructive remarks
that have improved the presentation of our paper.

\section*{Funding}
This work was supported in part by the Swedish Research Council (Vetenskapsr{\aa}det, 2019-04234),
the National Science Foundation under grant no.\ NSF AST-2307698 and a NASA ATP Award 80NSSC22K0825.
We acknowledge the inspiring atmosphere during the program on the
``Generation, evolution, and observations of cosmological magnetic fields''
at the Bernoulli Center in Lausanne.
We acknowledge the allocation of computing resources provided by the
Swedish National Allocations Committee at the Center for Parallel
Computers at the Royal Institute of Technology in Stockholm and
Link\"oping.

\section*{Declaration of Interests}
The authors report no conflict of interest.

\section*{Data availability statement}
The data that support the findings of this study are openly available
on Zenodo at doi:10.5281/zenodo.10527437 (v2024.11.02) or, for easier access,
at \url{http://norlx65.nordita.org/~brandenb/projects/Two-Conserved/}.
All calculations have been performed with the {\sc Pencil Code}
\citep{JOSS}; DOI:10.5281/zenodo.3961647.

\section*{Authors' ORCIDs}

\noindent
A. Brandenburg, https://orcid.org/0000-0002-7304-021X

\noindent
A. Banerjee, https://orcid.org/0009-0004-6288-0362

\bibliography{ref}{}
\bibliographystyle{jpp}
\end{document}